\documentclass[12pt]{iopart}

\usepackage{amsfonts,amssymb}
\usepackage{graphicx,epsfig}
\usepackage{bm}

\def\be{\begin{equation}}
\def\ee{\end{equation}}
\def\bea{\begin{eqnarray}}
\def\eea{\end{eqnarray}}
\def\bfa{\begin{mathletters}}
\def\ema{\end{mathletters}}
\def\P{{\cal P}}
\def\0{\overline{0}}

\def\q0{\underline{0}}

\def\S{{\cal S}}

\def\one{\leavevmode\hbox{\small1\normalsize\kern-.33em1}}

\def\ew#1{\langle#1\rangle}
\newcommand{\eins}{\mathbf{1}}

\newcommand{\aop}{\hat a}
\newcommand{\aopd}{\hat a^{\dagger}}

\begin{document}

\title{Quantum polarization spectroscopy of correlations in attractive fermionic gases}
\author{T. Roscilde$^1$, M. Rodr\'iguez$^2$, K. Eckert$^3$, O. Romero-Isart$^3$ , M. Lewenstein$^{2,4}$,
E. Polzik$^5$ and  A. Sanpera$^{3,4}$}
\address{$^1$ Laboratoire de Physique, Ecole Normale Sup\'erieure de Lyon, 46 All\'ee d'Italie,
69007 Lyon, France}
\address{$^2$ ICFO--Institut
de Ci\`encies Fot\`oniques, E-08860, Castelldefels, Spain}
\address{$3$ Grup de F\'isica Te{\`o}rica, Universitat Aut\`onoma
de Barcelona, E-08193, Bellaterra, Spain} \address{$^4$ ICREA--
Instituci\'o Catalana de Recerca i Estudis Avan\c cats, E-08010,
Barcelona, Spain}
\address{$^5$ Niels Bohr Institute, Danish Quantum Optics Center
QUANTOP, Copenhagen University, Copenhagen 2100, Dennmark}


\begin{abstract}
We show how spin-spin correlations, detected in a non-destructive
way via spatially resolved quantum polarization spectroscopy,
strongly characterize various phases realized in trapped ultracold
fermionic atoms. Polarization degrees of freedom of the light
couple to spatially resolved components of the atomic spin. In
this way quantum fluctuations of matter are faithfully mapped onto
those of light. In particular we demonstrate that quantum spin
polarization spectroscopy provides a direct method to detect the
Fulde-Ferrell-Larkin-Ovchinnikov phase realized in a
one-dimensional imbalanced Fermi system.
\end{abstract}

\section{Introduction}

Condensation of fermionic pairs occurs in nature in a variety of
systems, such as neutron stars, atomic nuclei, excitons in solids
and superconducting materials.  Ultracold Fermi gases have been in
the last years at the forefront of research both theoretically and
experimentally \cite{reviews}. Indeed homonuclear mixtures 
of fermionic atoms in two different hyperfine (pseudo-spin)
states offer the unprecedented advantage of a continuous tuning of
the interspecies attractive interaction. Thanks to this unique feature recent
experiments have spectacularly demonstrated high-temperature superfluidity 
of attractive fermions \cite{Jila,Grimm,Thomas,MITv} for a large
interval of values of the scattering length, spanning for the
first time the crossover from Bardeen-Cooper-Schrieffer (BCS) 
pairing to the Bose-Einstein condensation (BEC) of bosonic
molecules composed of two fermions (BCS-BEC crossover)
 \cite{Leggett}. The fate of the fermionic superfluid upon imbalancing 
 the two spin
species has been monitored in elongated traps, showing that
conventional pairing accompanied by segregation of the excess
majority atoms persists over a large interval of imbalance values
\cite{MIT,Rice}. This finding stands in contrast to the
expectation for one-dimensional systems, where exotic pairing with
finite-momentum pairs of Fulde-Ferrell-Larkin-Ovchinnikov (FFLO)
type \cite{FFLO} is predicted to occur
\cite{QMC,DMRG,Feiguinetal07,exactH,exactT}.
In addition, ultracold heteronuclear fermionic mixtures \cite{Willeetal08}
with controllable interactions have been very recently realized, as well  
Bose-Fermi mixtures in which the attractive interaction can create
a Fermi gas of polar molecules \cite{KKNi08}. 

 One of the major difficulties confronted by recent experiments
is the problem of faithful detection of correlations in the
strongly correlated regime of the gas. For attractive fermionic
gases, the best-developed probe for phase correlations to date is
represented by the dynamical projection of fermionic pairs onto
deeply bound molecules and the destructive time-of-flight
measurement of the latter \cite{Jila,MIT,Rice}. In the limit of
deeply bound molecules, noise correlation analysis
\cite{Altmanetal04} has been used as well to detect pairing
correlations \cite{Regaletal05}. The other crucial probes used in
experiments are phase-contrast imaging \cite{MIT,Rice} which
reveals only the local correlations between the two species;
imaging of the vortex lattice induced by stirring, which reveals
macroscopic phase coherence \cite{MITv}; and radio-frequency
spectroscopy \cite{Grimm, Schunketal08}, which probes the binding energy of the
fermion pairs.

 In this paper we focus on the fundamental insight that can be
gained by shining polarized light onto the atoms and detecting
the quantum fluctuations imprinted onto the light polarization
by the atomic sample (quantum polarization spectroscopy, QPS) \cite{QPS}.
This non-destructive measurement gives direct access to
\emph{spin-spin} correlations in the atomic system \cite{Eckertetal07}.
When the light shone on the atomic sample
is a standing wave \cite{Eckertetal08}, this type of measurement
allows to directly probe the magnetic structure factor at the wavevector
corresponding to the standing-wave period.
Here we focus on the case of trapped attractive fermions, and we show
that the measurement of the fluctuations of the light quadratures -- which
gives the information on spatially resolved spin correlations and fluctuations --
is very sensitive to various aspects of the paired phases
occurring in this system. In the case of a balanced Fermi gas
undergoing a crossover from the BCS regime to the BEC regime,
we show that the evolution of spin-spin fluctuations at a given wavevector
exhibits directly the shrinking of the size of the pairs
upon varying the scattering length. In
the case of a one-dimensional imbalanced Fermi gas in an optical lattice,
we show that FFLO pairing leaves an unambiguous fingerprint on
spin-spin correlations. This signature
enjoys the full robustness of the FFLO phase in 1D
\cite{QMC,DMRG, Feiguinetal07, exactH,exactT},
and it persists also in presence of a parabolic trapping
potential.

 The structure of the paper is as follows. In section
\ref{sec:2} we introduce the atom-light interfaces and the
interaction Hamiltonian and show how the atomic spin fluctuations
are mapped onto the fluctuations of light. Section \ref{sec:3}
deals with balanced Fermi mixtures, and the fate of spin-spin
correlations across the BEC-BCS crossover. Section \ref{sec:4}
focusses on an imbalanced Fermi mixture in a one-dimensional
optical lattice, and demonstrates the FFLO fingerprint on
spin-spin correlations. Conclusions are discussed in Section
\ref{sec:5}. Before proceeding further, let us stress that the
method we propose here to unveal spin-spin correlations in
ultracold atomic gases is not restricted to atomic
samples. It could be extended, e.g., to study the 
spin physics of fermionic molecules created in Bose-Fermi
mixtures. Fermionic molecules might posses very large electric
dipole moments which modifies substantially the properties of the
strongly interacting regime with respect to the case of short-range 
interactions. QPS could hence help to diagnose the
effects of dipole interactions on the spin correlations 
in such a regime.

\section{\label{sec:2} Atom-light interfaces}


A thorough derivation of the atom-light interafaces that result
from the propagation of polarized light in an atomic sample can be
found in \cite{julsgaard}. The effective dipole interaction between an atom with spin {\bf J} and a linearly polarized off-resonant light propagating in an arbitrary direction  reads
\begin{eqnarray}\!\!\!\!\!\!\!\!\!\!\!\!\!\!\!\!\!\!\!\!\!\!\!\!\!\!\!\!\!\!\!\!\!\!\!
H_{\rm int}^{\rm eff}=-a \int d{\bm r} \left[a_0 \hat
\eins_{ph}\hat \eins_{at}+ \hat s_z \hat J_z
-\frac1{\sqrt2}\left(\aopd_z\aop_++\aopd_-\aop_z\right)\hat J_+
-\frac1{\sqrt2}\left(\aopd_z\aop_-+\aopd_+\aop_z\right)\hat
J_-\right].\label{eqn:effHam}
\end{eqnarray}
Here, $\aop^{\dagger}({\bm r},t)$ and $\aop({\bm r},t)~$ denote
the creation and annihilation electric field operators, 
$\hat{s}_z({\bm r},t)=
\frac12\left(\aopd_+\aop_+-\aopd_-\aop_-\right)$ is the Stokes
operator, and
$\hat{\eins}_{at(ph)}$ denotes the total density of atoms
(photons). The constant coupling $a = a_0
c\gamma\lambda^2\hbar /(\mathcal{A}\delta\pi)$
 where $\gamma$ is the excited state linewidth, $\lambda$ the wavelength of the probing laser, $\mathcal{A}$ the cross section of the
probing laser overlapping with the atomic sample, $\delta$ is the
detuning and $a_0$ is the standard AC Stark shift (for quantum number
$F=3/2$, $a_0=3$). For a two-component Fermi system the atomic
spin $\hat{{\bm J}}$ can be defined
 in terms of the field operators
$\hat\psi_{\sigma =\uparrow\downarrow}^\dagger({\bm r},t)$ in the
usual manner
\begin{eqnarray}\!\!\!\!\!\!\!
\hat{J}_z=\frac{1}{2}(\hat\psi_\uparrow^\dagger\hat\psi_\uparrow-
\hat\psi_\downarrow^\dagger\hat\psi_\downarrow), \hspace{1cm}
\hat{J}_x=\frac{1}{2}(\hat
J_{+}+\hat J_{-}), \hspace{1cm}
\hat{J}_y=\frac{1}{2i}(\hat J_{+}-\hat
J_{-}),
\end{eqnarray}
where $\hat J_{+}({\bm r},t)=\hat\psi_\uparrow^\dagger
\hat\psi_\downarrow$, $\hat J_{-}({\bm
r},t)=\hat\psi_\downarrow^\dagger \hat\psi_\uparrow$.

\begin{figure}[t]
\begin{center}
\includegraphics[width=1.\linewidth]{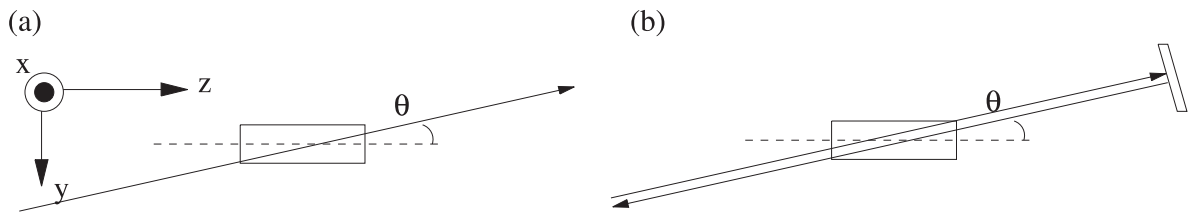}
\end{center}
\caption{ (a) Setup for a single beam with intensity $I({\bm r})$
propagating in direction $\theta$; (b) Two plane waves
counterpropagating in direction $\theta$, giving rise to a
standing wave configuration in the propagation direction. The
second one is obtained by a mirror reflection of the first one.}
\label{fig:setups}
\end{figure}

We consider the different probing configurations shown in
Fig.~\ref{fig:setups}. They lead to an effective interaction
Hamiltonian of the form \footnote{Here and in the following
equations, the coordinate labels $(x,y,z)$ of the Stokes operators
are associated with a reference frame in which 
the beam propagates along the $z$ direction.}
\begin{equation}
H_{\rm int}^{\rm eff}=- a\int d{\bm r}
A({\bm r})\hat s_z \hat J_P
\end{equation}
where $\hat J_P=\hat J_z\cos\theta-\hat J_y\sin\theta$ and the
intensity $A({\bm r})=I({\bm r})$ for a tilted beam shown in fig.~
\ref{fig:setups} (a). Probing setup (b) leads to $A({\bm
r})=2\cos^2[k r_P+\phi]$, where $r_P=z\cos\theta-y\sin\theta$ and
$\phi$ is the spatial phase of the probing standing wave. 

In general both the atomic spin vectors and the Stokes operators
are functions of space and time coordinates. Neglecting
retardation effects, we can integrate over the duration of the
probe pulse (of the order of $\mu s$) and define a macroscopic
Stokes operator $\hat S_\alpha =\int dt \hat s_\alpha$ where
$\alpha=x,y,z$. We consider light initially polarized in the $x$-
direction $\ew{\hat S_x}=N_P/2$ and $\ew{\hat S_y}=\ew{\hat
S_z}=0$, where $N_P$ is the photon number. Heisenberg equations of
motion for the atomic spin lead to $\hat J_\alpha({\bm r},t)\equiv
\hat J_\alpha({\bm r})$ $\forall \alpha$ to first order in time,
while the spin in the direction of the probe $\hat J_P$ is
conserved to all orders. Due to the atom-light interaction, the
Stokes operator performs a Faraday rotation in the plane
perpendicular to propagation
\begin{eqnarray}
\partial_{\bm r} \hat S_y = - a \hat J_P ({\bm r})\frac{N_P}{2}A({\bm
r}). \label{eq:syc}
\end{eqnarray}
Integrating equation (\ref{eq:syc}), we find that the light
quadrature, defined by the Stokes operators $\hat S_y$, $\hat
X_S=\sqrt{2/N_P}\hat S_y$  reads after propagation through the
sample
\begin{equation}
\ew{\hat X_s^{\rm out}}=\ew{\hat X_s^{\rm
in}}-\frac{\kappa}{\sqrt{N_A}}\int d {\bm r} \ew{\hat J_P({\bm
r})}A({\bm r}),
\end{equation}
where $\kappa^2=a^2N_PN_A/2$, being $N_A$ the total number of atoms. The fluctuations read
\begin{eqnarray}
&&( \delta \hat X_S^{\rm out})^2=\frac12+
\frac{\kappa^2}{N_A}\int d{\bm r} A({\bm r})\int d{\bm r'} A({\bm
r'}) \ew{\delta \hat J_P ({\bm r}) \delta \hat J_P({\bm r'})},\label{eq:fluc}
\end{eqnarray}
where $\delta \hat J_P = \hat J_P - \langle \hat J_P \rangle$.
Thus fluctuations of the light quadrature (polarization
fluctuations) after crossing the atomic sample contain the photon
shot noise plus a contribution proportional to second order
correlations of the atomic spins. Experimentally,  spin-spin
correlations can be detected if their contribution in the
polarization fluctuations is larger that the photon shot noise,
which for a coherent initial source corresponds to $( \delta \hat
X_S^{\rm in})^2=\frac12$. To this end, the relevant parameter is
$\kappa^2= \eta \alpha$ where $\alpha$ is the {\it resonant}
optical depth of the sample and $\eta$ is the spontaneous emission
probability \cite{QPS,Eckertetal08}. It can be shown that the
optimal signal is obtained when $\eta$ is tuned to
$\eta_{opt}=1/\sqrt{2\alpha}$ and $\kappa^2=\sqrt{\alpha/2}$
\cite{kappa}. BEC clouds have typical optical depths of
the order of a few hundreds, for which one obtains quantum fluctuations
imprinted on light by the spin fluctuations which are
significantly bigger than the photon shot noise.

Technically it might be challenging to fix the spatial phase of
the probing light $\phi$ with respect to the trapping potential of
the atoms in standing wave probing configuration (b). We assume
that the average of the signal over various shots of the
experiment leads to averaging over $\phi$. Under phase averaging,
Eq.~(\ref{eq:fluc}) reduces to
\begin{eqnarray}
&&( \delta \hat X_S^{\rm out}(\bm k))^2=\frac12+ \frac{\kappa^2
V}{4 N_A} \left[ 4 S_m(0) +S_m(2 \bm k)+S_m(-2\bm k) \right]  .
\label{eq:phav}
\end{eqnarray}
Here we have introduced the magnetic structure factor
\begin{equation}
S_m(\bm k) = \frac{1}{V}\int d\bm r \int d \bm r' e^{i \bm k (\bm r-\bm r')} \ew{ \delta \hat J_P ({\bm r}) ~\delta \hat J_P({\bm r'})}
\label{eq:smk}
\end{equation}
where $V$ is the volume of the system and $\bm k$ is the momentum of the probing standing wave.
If the Hamiltonian of the system conserves the total spin along the probing direction $P$,
as it will be the case in the following examples, then $S_m(0)=0$. Moreover, due to the 
commutation of $\delta \hat{J}_P({\bm r}$ at different locations, $S(-2\bm k) = S(2\bm k)$.

In atoms are trapped in a one dimensional optical lattice
as considered in Section \ref{sec:4}, the
atomic spin field ($e.g.$ along the $z$ direction) is localized at
lattice sites, and it is conveniently expressed as
\begin{equation}
\hat{J}_z ({\bm r}) = \frac{1}{2} \sum_i |w({\bm r}-{\bm r}_i)|^2 ~\hat m_i
\end{equation}
where $\sum_i$ runs over the sites of the optical lattice
and $\hat m_i= (\hat{n}_{i\uparrow}-\hat{n}_{i\downarrow})/2$ where $\hat n_{i,\sigma}$ is the occupation of fermions with spin $\sigma$
at lattice site $i$. $w({\bm r}-{\bm r}_i)$ is the Wannier
function at lattice site $i$; in the following we will
approximate it with a $\delta$-function for simplicity.
If the probing standing wave is not phase-locked spatially to the
one creating the optical lattice, averaging over the relative
phase between the two leads to Eq.~(\ref{eq:phav}) which contains the lattice magnetic structure factor:
\begin{equation}
S_m({\bm k})=\frac{|\tilde{w}({\bm k})|^4}{L^d}\sum_{ij}e^{i \bm k \cdot (\bm r_i- \bm r_j)} \left(\ew{\hat m_i \hat
m_j}-\ew{\hat m_i} \ew{\hat m_j}\right).\label{eq:ssf}
\end{equation}
where $\tilde{w}$ is the Fourier transform of the Wannier function, and $L$ is the 
linear dimension of the $d$-dimensional lattice. In the following we
consider for simplicity $w({\bm r}) \sim \delta({\bm r})$, and consequently we neglect
the ${\bm k}$-dependence of  $\tilde{w}$. 

Hence the method proposed here gives direct access to the magnetic structure factor of the atomic sample.
This piece of information is crucial in detecting the onset of spin-spin correlations in the
strongly interacting phases of trapped atomic samples.
The (pseudo)spin degree of freedom of spin-$S$ atoms
is encoded in $2S+1$ internal hyperfine states which are populated in the atomic sample.
A fundamental example of spin correlations is the antiferromagnetic
phase emerging at low temperatures in a  Mott insulator of spin-(1/2) bosons or fermions 
in an optical lattice \cite{KuklovS03, Demleretal03}. 
More exotic spin states appear in spin-1 bosons, including dimerized states
in one dimensional optical lattices \cite{Rizzietal05}.
Moreover, recent proposals envision the dipolar coupling of a $S=1/2$ pseudospin degree of
freedom of molecules dressed with microwaves \cite{Michelietal06}: the spin couplings realized in this
way give rise to novel magnetic phases including, \emph{e.g.},  topological quantum order.

\section{\label{sec:3} Three dimensional balanced Fermionic superfluids}
We begin by considering superfluidity in an homogeneous two-component balanced 3D
Fermi system with two-body contact attractive
interactions. The presence of interactions leads to pairing between the different spins described by
the pairing order parameter $ \Delta({\bm r})= |g|\ew{\hat
\psi^\dagger_\uparrow({\bm r})\hat \psi^\dagger_\downarrow({\bm
r})}$,  where the interaction strength $g \propto 2k_F a_s$,  $a_s$
is the two-body scattering length, and $k_F$ is the Fermi vector.
For an homogeneous balanced 3D gas $\ew{\hat {\bm J}({\bm r})}=0$
and the spin fluctuations are the same along all directions. Using
Wick's theorem, the fluctuations in $\hat J_z$ read
\begin{equation}
\hspace{-2.5cm}
4\ew{\delta \hat J_z ({\bm r})\delta \hat J_z({\bm r'})}=
\sum_{\sigma=\uparrow\downarrow} ( \ew{\hat n_\alpha ({\bm
r})}\delta ({\bm{r-r'}}) -|\ew{\hat \psi_\sigma^\dagger({\bm r})
\psi_\sigma({\bm r'})}|^2 ) -2 |\ew{\psi^\dagger_\uparrow({\bm
r})\psi^\dagger_\downarrow({\bm r'})}|^2 .
\end{equation}
The BCS formalism \cite{Leggett} describes the BCS-BEC crossover,
characterized by different values of the scattering length
$1/k_Fa_s$, in terms of the Bogoliubov amplitudes
$u_{\textbf{k}}^2(v_{\textbf{k}}^2)=\frac12\left(1\pm\frac{\textbf{k}^2-\mu}{\sqrt{\Delta^2+(\textbf{k}^2-\mu)^2}}\right)$
where the chemical potential $\mu$ and superfluid gap $\Delta$ are
obtained from the simultaneous solution of the gap and number
equations \cite{crossover}:
\begin{eqnarray}
& \frac{1}{k_Fa}=(\tilde \mu^2+\tilde \Delta^2)^{1/4}P_{1/2}(x) \label{gap} \\
& \frac{\pi}{4}=\tilde\mu (\tilde \mu^2+\tilde
\Delta^2)^{1/4}P_{1/2}(x) +(\tilde \mu^2+\tilde
\Delta^2)^{3/4}P_{-1/2}(x) \label{num}
\end{eqnarray}
where $x=-\mu/(\mu^2+\Delta^2)^{1/2}$, $\tilde
\Delta=\Delta/\epsilon_F$, $\tilde \mu=\mu/\epsilon_F$ and
$P_\eta(x)$ are the Legendre functions of the first kind.

We consider two types of probing configurations: a running
Gaussian beam [Fig.~\ref{fig:setups}(a)] and a phase-averaged
standing-wave configuration [Fig.~\ref{fig:setups}(b)]. A Gaussian
probing beam has an amplitude profile $A({\bm r})=e^{-{|{\bm
r}_{\perp}|^2}/{\sigma^2}}$, where ${\bm r}_{\perp}$ are the
coordinates perpendicular to the propagation direction. When
the light is macroscopically polarized in the $x$-direction propagating through
the sample, it yields $\ew{\hat X_s^{\rm out}}=\ew{\hat X_s^{\rm
in}}$ and
\begin{eqnarray}
(\delta X^{\rm out}_s({\bf
k}))^2-\frac{1}{2}=\frac{\kappa^2}{2}\left[ 1- \frac{V}{N_A}\int
d{\bm r}~e^{-r_\perp^2/\sigma^2} \left( |I_{\uparrow \uparrow}(\bm
r)|^2+ |I_{\downarrow \uparrow}(\bm r)|^2  \right) \right],
\end{eqnarray}
where
\begin{eqnarray}
I_{\uparrow \uparrow}(\bm r)=\frac{1}{(2\pi)^2}\int_0^\infty d k~ k^2
~v_{\bf k}^2 ~j_0(k r) \nonumber \\
I_{\uparrow \downarrow}(\bm r)=\frac{1}{(2\pi)^2}\int_0^\infty d
k~ k^2 ~v_{\bf k} u^*_{\bf k} ~j_0(k r),
\end{eqnarray}
and $j_0$ is the Bessel function.

For a homogeneous probing set-up the total spin fluctuations are
zero \cite{cambridge}, while a finite Gaussian probe shows a
finite value of the spin fluctuations.
\begin{figure}[t]
\begin{center}
\includegraphics[width=0.7\linewidth]{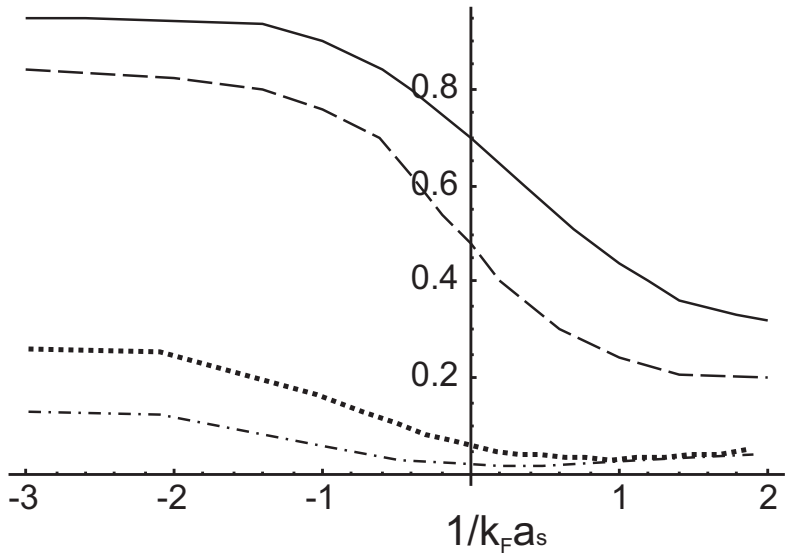}
\end{center}
\caption{ Spin fluctuations $((\delta X^{\rm
out}_s)^2-\frac{1}{2})/\frac{\kappa^2}{2}$ in the BCS ($a<0$)-BEC
crossover for a Gaussian probing beam. Sizes $k_F\sigma=0.5$
(solid) , $k_F\sigma=1$ (dashed), $k_F\sigma=5$ (dotted) and
$k_F\sigma=10$ (dot-dashed). Here $V=L^3$ with $Lk_F=100$.}
\label{fig:gaus}
\end{figure}

\begin{figure}[t]
\begin{center}
\includegraphics[width=0.9\linewidth]{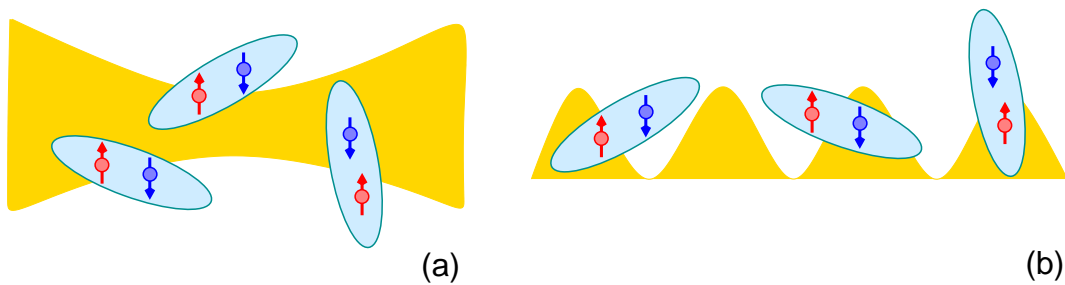}
\end{center}
\caption{Probing schemes of spin fluctuations in paired fermionic gases. (a) Gaussian laser beam.
(b) Standing wave.} \label{fig:probe}
\end{figure}

Results in Fig. \ref{fig:gaus} show how the polarization
fluctuations decrease with decreasing pair size along the BCS-BEC
crossover. These results are in agreement with those of
Ref.~\cite{bruun},  and they reveal that the quantum fluctuations
of the atomic spins imprinted on the light polarization are
significantly suppressed when the characteristic pair size becomes
smaller than the Gaussian beam waist $\sigma$. This is due to the
simple fact that an $s$-wave pair is in a total spin singlet, and
hence it does not contribute to the spin fluctuations of the atoms
illuminated by the laser beam when it is fully contained within
the beam waist. A sizable quantum noise imprinted in the
polarization corresponds to the optimal situation in which the
pair size is larger than or comparable to the beam waist, as
sketched in Fig.~\ref{fig:probe}(a). Yet a fundamental remark is
necessary: in order to obtain a total excursion of order ${\cal
O}(1)$ in the noise signal along the crossover, one needs to focus
the laser to the experimentally challenging waists $\sigma \sim
1-10 ~  k_F^{-1}$. In fact, for typical sample densities of order
$n\sim 10^{13}$ atoms/cm$^3$, one has
 $k_F^{-1} = (3 \pi^2 n)^{-1/3} \sim 150$ nm.

\begin{figure}[t]
\begin{center}
\includegraphics[width=0.7\linewidth]{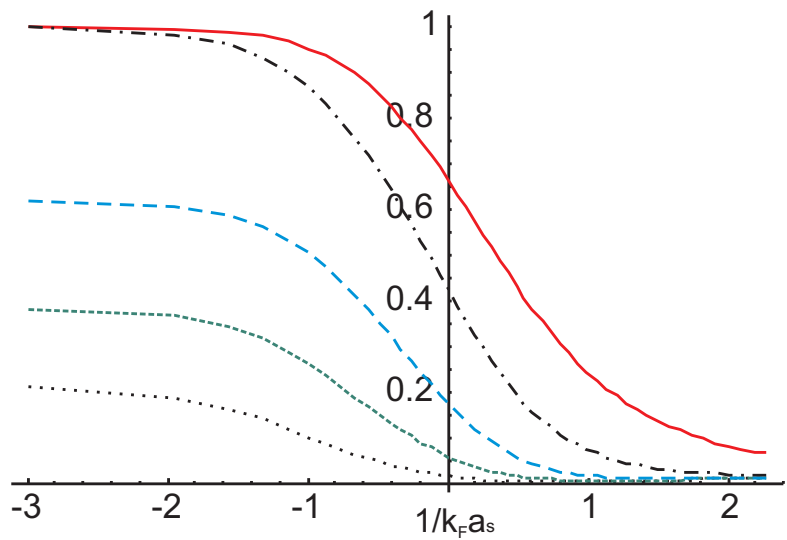}
\end{center}
\caption{ Spin fluctuations $((\delta X^{\rm
out}_s)^2-\frac{1}{2})/\frac{\kappa^2}{2}$ for different momenta
of a phase-averaged probing standing wave $k$ in the BCS
($a<0$)-BEC crossover. This is also proportional to the magnetic
structure factor $S_m(2 k)$. Different lines correspond to
$k/k_F=0.25,0.5,1,2,3$ in ascending order.} \label{fig:struc}
\end{figure}

On the other hand, probing the system with a standing wave, as
proposed in the previous section, gives high spatial resolution
without the need of focusing a laser over prohibitively small
length scales.
 Fig.~\ref{fig:struc} shows the
 magnetic structure factor $S_m(2 k) V/N_A$, Eq.~(\ref{eq:smk}),  at fixed wavevector $k$, obtained via the phase-averaged polarization fluctuations
 of two counter-propagating beams at wavevector $k$,
 as a function of the product $(k_F a_s)^{-1}$.  Strongly correlated spin fluctuations,
 recorded by the magnetic structure factor, are present over length scales related to the pair 
 size, while the inter-pair spin correlations are vanishing. Hence the
 magnetic structure factor is very strongly affected by the shrinking of the pairs controlled
 by the scattering length, which gradually pushes the
 correlation length of spin fluctuations below the probe wavelength $2\pi/k$ (Fig.~\ref{fig:probe}(b)).
 The light wavevectors $k/k_F$ = 0.25, 0.5, 1, 2, and 3, considered
 in Fig.~\ref{fig:struc}, correspond
 to wavelengths $\sim$ 3770, 1885, 942,  471, and 314 nm respectively. The strongest excursion
 on the magnetic structure factor along the BCS-BEC crossover is observed for the largest wavevectors,
 which correspond to conventional laser wavelengths. 
 The other wavelengths can be obtained
 effectively by crossing the counter-propagating beams at an angle, as we will discuss
 in a forthcoming publication.

\begin{figure}[t]
\begin{center}
\includegraphics[width=0.55\linewidth]{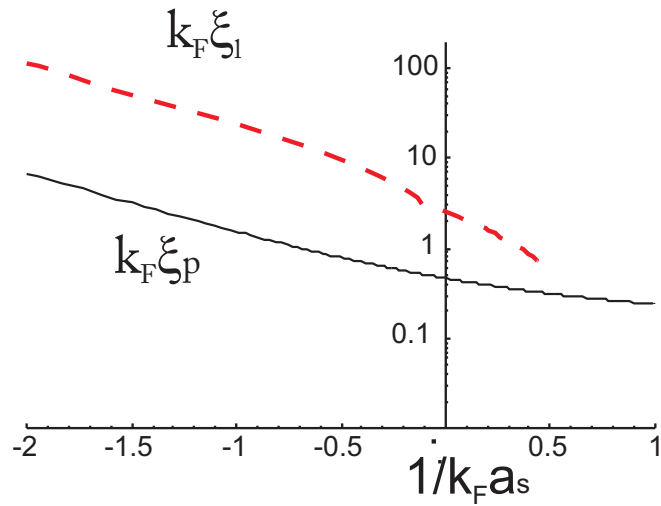}
\end{center}
\caption{Solid line shows the Pippard coherent length
($\xi_p=1/\Delta \pi$) in $\log$-scale while the dashed line shows
the corresponding length $\xi_l=2\pi/k_0$ where $k_0$ is the
inflection point of the magnetic structure factor $S_m(2 k)$. }
\label{fig:pairsize}
\end{figure}

  The inflection point of the magnetic structure factor $S_m(2k)$ as a function of $(k_F a)^{-1}$
 contains the information about the characteristic length $\xi_l$ beyond which  
 the spatial average of spin fluctuations imprinted on the light polarization 
is strongly suppressed. The various scans of $S_m(2k)$ at fixed $k$
for varying  $(k_F a)^{-1}$ allow to extract the $(k_F a)^{-1}$ location
of the inflection point, and to reconstruct how the associated wavelength
$\xi_l = 2\pi/k$ depends upon $(k_F a)^{-1}$. This dependence is shown 
in Fig.~\ref{fig:pairsize}. It is remarkable to see that the length $\xi_l$ 
follows the same behavior of the Pippard coherence length $\xi_p$ (up to a scaling factor) 
on the BCS side of the crossover, while it starts decreasing much faster
around the unitarity limit. In this respect, it would be desirable to go beyond
mean-field theory in the description of the strongly interacting regime, and to extend
to that regime the comparison of the spatial features contained in the structure
factor with the characteristic pair size. 
 
 The above analysis can be applied to BCS pairs as well as to
 molecules. Therefore QPS based on standing waves
 appears as a promising method to probe the internal structure of $s$-wave Feshbach molecules.
 Further analysis of the QPS signal in the case of, e.g., $p$-wave or $d$-wave pairing
 and Feshbach molecules will be the subject of future investigations.

\section{\label{sec:4} Imbalanced Fermions in one dimension. Fingerprint
of the FFLO phase}

 The possibility of exotic pairing in imbalanced fermionic mixtures
has been recently the subject of a very intense research. In particular
the mismatch between the Fermi momenta $k_{F\uparrow}$ and
$k_{F\downarrow}$ of the two spin species
can lead to the appearance of FFLO pairs with finite momentum
$Q = |k_{F\uparrow} - k_{F\downarrow}|$.
The experiments carried out so far seem to rule out this
possibility in three dimensions, where phase separation
of the gas into a balanced mixture (with conventional
pairing) and a fully polarized
gas of the remaining majority atoms is observed \cite{MIT,Rice}.
On the other hand, analytical and numerical studies have
rigorously proved the occurrence of a highly stable
FFLO phase in one dimension \cite{QMC, DMRG, Feiguinetal07, exactH, exactT}.

In the following we concentrate on the case of one-dimensional
imbalanced attractive fermions in an optical lattice,
described by the 1D attractive Hubbard model in a parabolic trap:

\begin{equation}
H=-t \sum_{i,\sigma}\left( \hat{c}_{i,\sigma}^\dagger
\hat{c}_{i+1,\sigma} +\rm{h.c.} \right)-U\sum_i  \hat{n}_{i\uparrow}
\hat{n}_{i\downarrow} \label{eq:Hub}
\end{equation}

\noindent where $ \hat{c}_{i,\sigma}^\dagger$ creates a fermion with spin
$\sigma=\uparrow\downarrow$ at site $i$ , $
\hat{n}_{i\sigma}= \hat{c}_{i,\sigma}^\dagger \hat{c}_{i,\sigma}$, $t$ is
the hopping matrix and $U$ is the on-site attractive interaction.

 We investigate this model by means of numerically exact
quantum Monte Carlo simulations based on the Stochastic Series
Expansion algorithm \cite{SSE} in the canonical ensemble \cite{Roscilde08}.
The temperature is set to reproduce the $T=0$ properties of the system.
As pointed out in Section \ref{sec:2} the method of quantum
polarization spectroscopy gives access to the magnetic structure
factor Eq.~(\ref{eq:ssf}), which is the quantity we calculate
numerically with quantum Monte Carlo.

It is well known that, for a bipartite lattice, the attractive
Hubbard model can
be mapped onto the repulsive one via a particle-hole
transformation on one of the two species,
$\hat{c}_{i\uparrow}^{\dagger} = (-1)^i ~
\hat{c'}_{i\uparrow}$, $\hat{c}_{i\downarrow}^{\dagger} =
\hat{c'}^{\dagger}_{i\downarrow}$. The fillings of the
$c'$ fermions are $n'_{\uparrow} = 1-n_{\uparrow}$
and $n'_{\downarrow} = n_{\downarrow}$, and accordingly
the Fermi wavevectors are transformed as
$k'_{F\uparrow} = \pi-k_{F\uparrow}$ and
$k'_{F\downarrow} = k_{F\downarrow}$.
In particular the spin-spin correlation function of the
attractive model maps onto the density-density correlation
function of the repulsive one
\begin{equation}
C_{mm}(r) = \langle \hat{m}_i \hat{m}_{i+r} \rangle -
\langle \hat{m}_i \rangle \langle \hat{m}_{i+r} \rangle=
\frac{1}{4} \left(\langle \hat{n}'_i \hat{n}'_{i+r} \rangle - n'^2 \right)
\end{equation}
where $\hat{n}'_i = \hat{n}'_{i\uparrow} + \hat{n}'_{i\downarrow}$.
The density-density correlation function for the repulsive Hubbard
model has been investigated extensively in the past via Bethe Ansatz,
and its general form is of the type
\begin{equation}
C'_{nn}(r) = \langle \hat{n}'_i \hat{n}'_{i+r} \rangle - n'^2 =
\sum_l B_l \frac{\cos(\alpha_l k'_{F\uparrow} r + \beta_l k'_{F\downarrow} r)}{r^{\Delta_l}}
\end{equation}
where the various terms appearing in the sum are imposed by selection rules
\cite{Essleretal05}. In particular it is found that
$(\alpha_l, \beta_l) = (2,0), (0,2), (2,2), ...$. To the best of
our knowledge, the amplitudes $A_l$ of the various contributions have not
been determined for the case of a spin-imbalanced Hubbard model at an arbitrary
filling.

\begin{figure}[t]
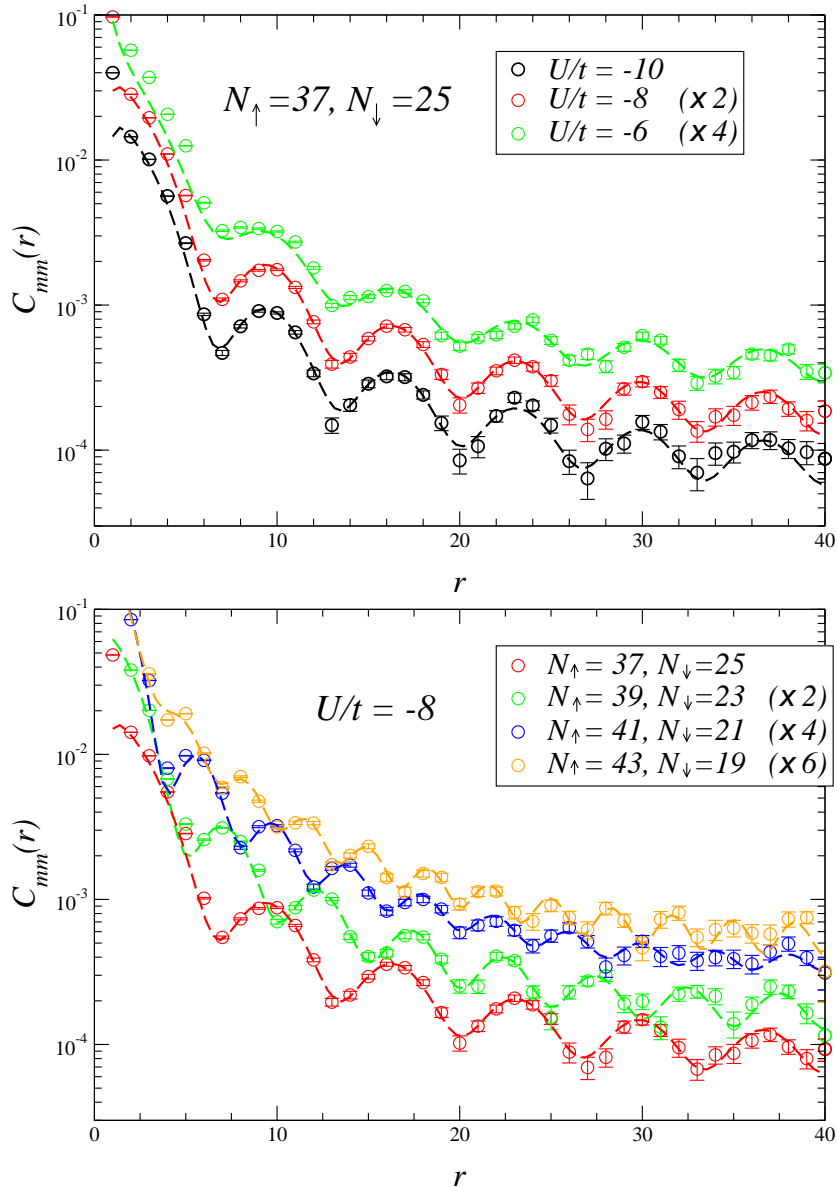

\begin{center}
\includegraphics[width=0.7\linewidth]{Cr-FFLO.eps}
\includegraphics[width=0.7\linewidth]{Cr-FFLO-Nscan.eps}
\end{center}
\caption{Spin-spin correlations in the 1D attractive Hubbard model
with population imbalance for different polarization
values and interaction strengths. The simulation data refer
to a system size of $L=80$. To improve
readability, in both panels some sets are multiplied by a constant
(indicated in the set legends). The dashed lines are fits
to the symmetrized form of Eq.~(\ref{eq:Cmm}). All results
show consistency with the estimates $\Delta_1 \approx 1.8$
and $\Delta_2 \approx 2-2.2$, although the precision on the
fitting coefficients does not allow us to extract the
systematic dependence of the exponents on the polarization
and interaction parameters.} \label{fig:spinspin}
\end{figure}

The spin-spin correlation function for the attractive imbalanced
1D Hubbard model is shown in Fig.~\ref{fig:spinspin} for
various values of the attraction and of the imbalance.
Our numerical findings are all consistent with the expression
\begin{equation}
\hspace{-2cm}
C_{mm}(r) = \frac{1}{4} ~C'_{nn}(r)
\approx A_1 ~\frac{1}{r^{\Delta_1}} + A_2 ~
\frac{\cos(2 k'_{F\uparrow} r + k'_{F\downarrow} r)}{r^{\Delta_2}} =
A_1 ~\frac{1}{r^{\Delta_1}}  + A_2 ~\frac{\cos(2Qr)}{r^{\Delta_2}}, ~~
\label{eq:Cmm}
\end{equation}
which, in the symmetrized form $C_{mm}(r) + C_{mm}(L-r)$,
provides excellent fits to the finite-size numerical data.

\begin{figure}[t]
\begin{center}
\includegraphics[width=0.7\linewidth]{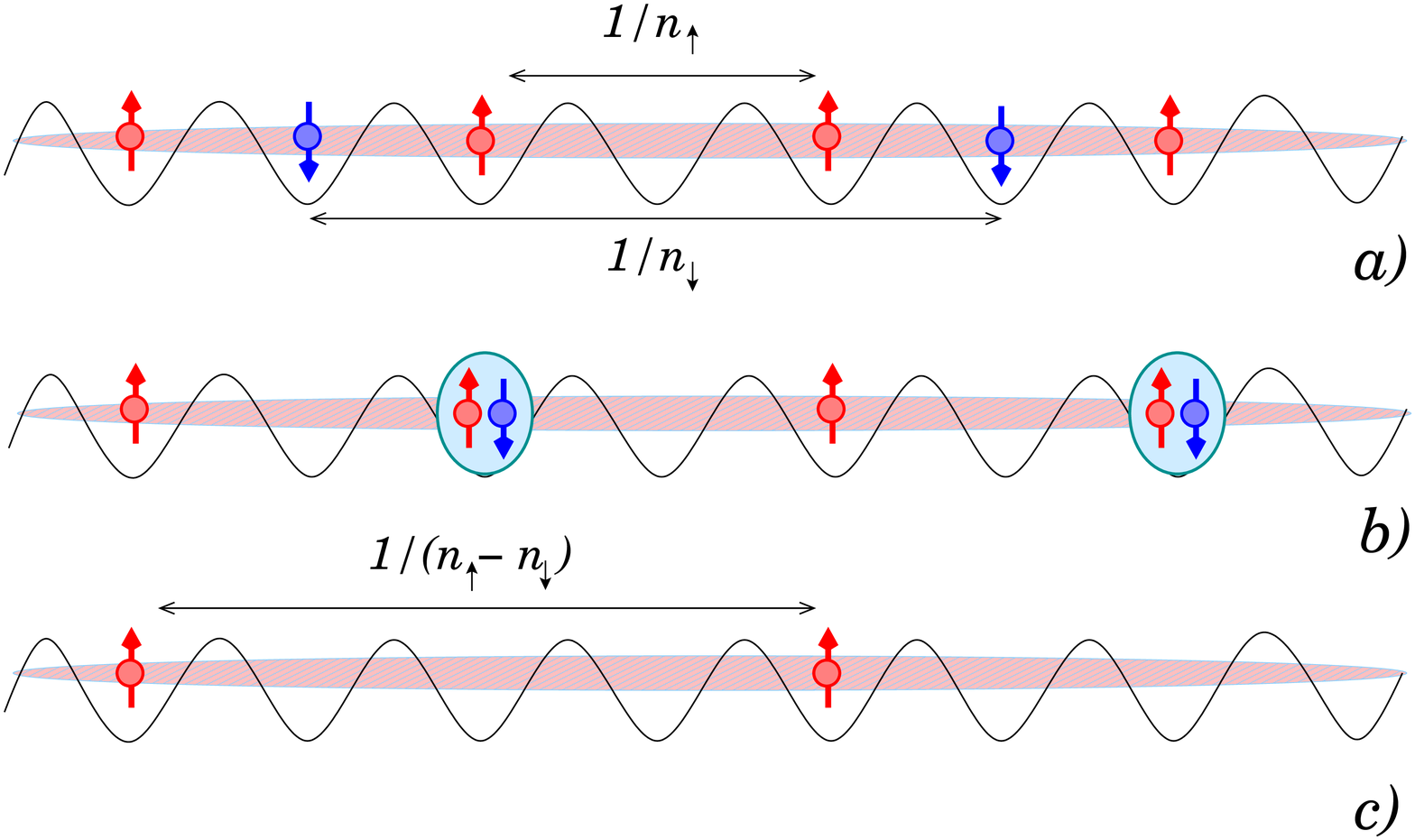}
\end{center}
\caption{Sketch of the short-range properties of an imbalanced
mixture of strongly attractive 1D fermions. $a)$ The two-component
gas displays strong fluctuations towards a short-range
crystalline-like arrangement with the majority spins at a maximum mutual
distance $1/n_{\uparrow}$, and similarly for the minority spins.
$b)$ The strongly attractive interaction leads to pairing fluctuations,
which set the two short-range crystalline arrangement in phase
spatially. From the magnetic point of view, pairing effectively
erases some of the itinerant spins in the system, so that
the resulting magnetic texture at short range has a characteristic
length $\approx (n_{\uparrow}-n_{\uparrow})^{-1}$ and a
characteristic associated wavevector $2\pi (n_{\uparrow}-n_{\uparrow}) = 2Q$.}
\label{fig:fflo}
\end{figure}

This shows that the spin-spin correlation function of the
attractive model is directly
sensitive to the Fermi momentum mismatch $Q$ at which FFLO pairing
occurs. A sketchy picture on the relationship between
spin-spin correlations and FFLO pairing in the strong pairing
limit is offered in Fig.~\ref{fig:fflo}. The FFLO state
is effectively a mixture of bound pairs and of unbound
majority atoms, which are mutually hardcore repulsive
due to Pauli exclusion principle. Infinite repulsion in
1D leads to algebraically decaying charge-density-wave (CDW)
correlations,
namely the system displays strong fluctuations towards
a local CDW state which very roughly corresponds to
an equally spaced arrangement of pairs and excess $\uparrow$
particles \cite{note.density}. From the point of view
of the spin texture, considering bound pairs as spinless
objects leaves out an algebraically decaying spin-density-wave
arrangement which has a characteristic wavevector
$2\pi (n_{\uparrow} - n_{\downarrow}) = 2Q$. Hence this
argument shows that the $2Q$-modulation of spin-spin
correlations is a direct fingerprint of pairing in an
imbalanced mixture.

\begin{figure}[t]
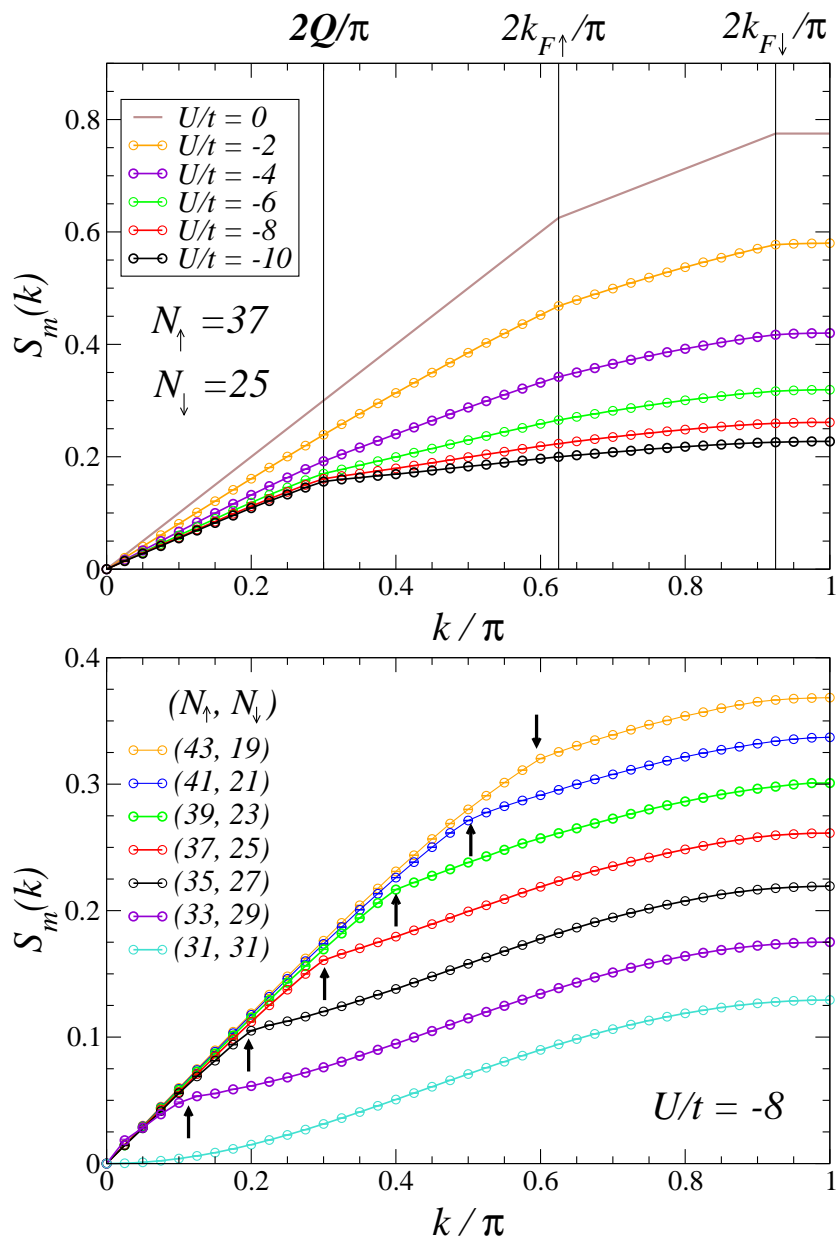

\begin{center}
\includegraphics[width=0.7\linewidth]{SzSzk12-N37-25L80.eps}
\includegraphics[width=0.7\linewidth]{SzSzk12-U8Nscan-L80.eps}
\end{center}
\caption{Magnetic structure factor in the 1D attractive Hubbard model
with population imbalance for different polarization
values and interaction strengths. The arrows in the lower
panel mark the location $2Q/\pi$.} \label{fig:Skm}
\end{figure}

 The $2Q$ modulation of spin-spin correlations translates
into a pronounced \emph{kink} at $k=2Q$ in the magnetic structure factor,
as shown in Fig.~\ref{fig:Skm} for various values of the
attraction and of the imbalance. This kink is shown to be
the only relevant feature in $S_m(k)$, and to be most pronounced
for small imbalance. Fig.~\ref{fig:Skm} also shows a comparison
with the case of two non-interacting spin species $U=0$, which
is exactly solvable. In that case the magnetic structure factor
reads (for $k_{F\uparrow} > k_{F\downarrow}$)
\begin{eqnarray}
S_m(k) &=& 
\frac{|k|}{\pi} ~~~~~~~~~~~~~ {\rm for}~~|k| \leq 2k_{F\downarrow}  \nonumber \\
&=& \frac{|k| + 2 k_{F\downarrow}}{2\pi}  ~~~~{\rm for}~~
2k_{F\downarrow} \leq |k| \leq 2k_{F\uparrow} \nonumber \\
&=& \frac{k_{F\uparrow} + k_{F\downarrow}}{\pi} ~~~~~ {\rm for}~~
2k_{F\uparrow} \leq |k| \leq \pi~~.
\end{eqnarray}
This expression clearly exhibits two independent kinks at $2k_{F\uparrow}$
and $2k_{F\downarrow}$. As shown in Fig.~\ref{fig:Skm}, upon increasing the
attraction $|U|$ among the two species, the two kinks of the non-interacting
case disappear gradually, while the kink at $2Q$ appears clearly for
$|U|/t \gtrsim 4$. The occurrence of a single kink at $2Q$ in the
interacting model is hence a direct consequence of pairing.

\begin{figure}[t]
\begin{center}
\includegraphics[width=0.8\linewidth]{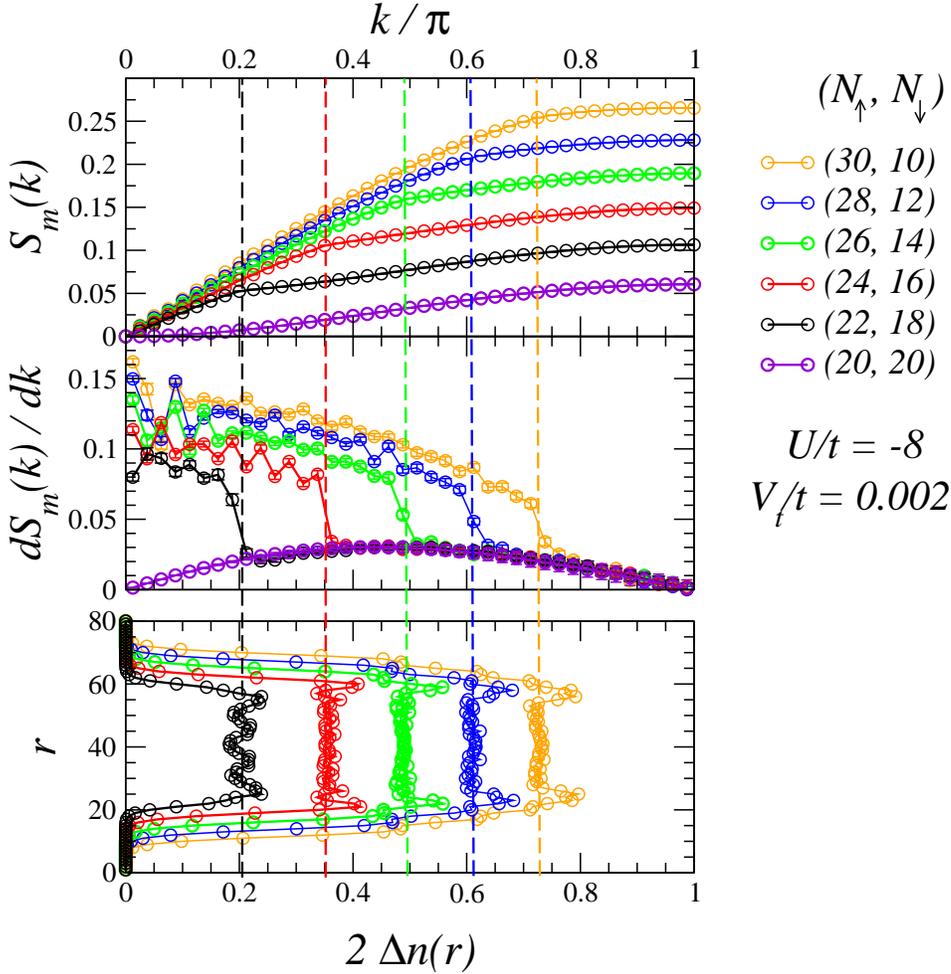}
\end{center}
\caption{\emph{Upper panel}. Magnetic structure factor in the \emph{trapped}
1D attractive Hubbard model with population imbalance for different polarization
values and interaction strengths.\emph{Middle panel}. Momentum derivative of the magnetic
structure factor. \emph{Lower panel}. Corresponding density
imbalance profiles, $\Delta n = n_{\uparrow} - n_{\downarrow}$.
The dashed lines mark the correspondence between the central density imbalance
in the trap, the position of the kink in the structure factor, and the corresponding
jump in the structure factor derivative.} \label{fig:trap}
\end{figure}

 To make full contact with experiments, we also add a parabolic
trap to the Hamiltonian Eq.~(\ref{eq:Hub})
\begin{equation}
{\cal H'} = {\cal H} + V_t \sum_{i,\sigma} (i-L/2)^2~n_{i,\sigma},
\end{equation}
where $L$ is the size of the lattice. We use similar parameters
to those of Ref.~\cite{Feiguinetal07}, which shows clear
evidence of FFLO pairing in terms of off-diagonal correlators.
Fig.~\ref{fig:trap} remarkably shows that
the kink feature in the magnetic structure factor survives to the presence of a
trap. In particular, the presence of a kink is even better evidenced by
the momentum derivative $dS_m(k)/dk$, which correspondingly exhibits
a marked jump. The persistence of this feature in a trap is intimately connected
with the fundamental fact that
the density imbalance $\Delta n = n_{\uparrow} - n_{\downarrow}$ is almost
constant over a significant portion of the trap center (although the
density profiles of the two species independently are not as flat).
Given that the location of the kink is only sensitive to the imbalance
$Q = \pi\Delta n$, the persistence of the same $\Delta n$ over a large
portion of the cloud protects the kink from smearing.
Conversely, the position of the kink can be regarded as an efficient
measure of the density imbalance in the trap center, which is not
in principle known a priori. This is a valuable alternative
to the direct \emph{in-situ} phase contrast imaging
recently applied to polarized Fermi gases \cite{Shinetal06}
for the measurement of the density profiles of both species.

\section{\label{sec:5} Conclusions}
We have shown how spin-spin correlations of an attractive Fermi gas can be detected in a non-destructive way using spatially resolved quantum polarization spectroscopy (QPS). The atomic spins couple to the polarization degree of freedom of light and they imprint their correlations on the quantum fluctuations of the polarization, which can be measured using homodyne detection. When a standing wave is shone on the atomic sample, the measured
signal allows a high resolution of spin-spin correlations in momentum space at the wavector of the
standing wave.

Spin-spin correlations are shown to strongly characterize the superfluid phases of fermionic
systems with attractive interactions. In a three-dimensional spin-balanced system the spatial structure of
spin-spin correlations depends strongly on the pair size and hence it evolves strongly along the crossover
from the BCS to the BEC regime: as a consequence, QPS is able to reveal the evolution of the characteristic pair size as a function of the scattering
length.  In a one-dimensional system with spin imbalance, the magnetic structure factor,
recorded by the QPS signal, shows a kink at the difference
between the Fermi vectors of the two spin species, providing a direct signature of
finite-momentum Fulde-Ferrell-Larkin-Ovchinnikov pairing.
 The proposed experimental technique is most promising to detect the spin structure of exotic
pairs and molecules, including \emph{e.g.} $p$-wave and $d$-wave pairing, and to detect the magnetic
phases which can be potentially realized by strongly correlated atoms and molecules loaded
in optical lattices.

\section{\label{sec:6} Acknowledgements}

We acknowledge the
support of the the European
Commission through the``SCALA" project and STREP ``NAMEQUAM".
the grant EMALI and FET grants
HIDEAS, number FP7-ICT-221906 and COMPAS.
Moreover we acknowledge the support coming from the grants of the 
Spanish Ministerio de Educaci\'on y
Ciencia (FIS2005-0462,  FIS2008-01236, Consolider Ingenio 2010 ``QOIT"),
the Catalan Goverment grant SGR00185,
and the ESF-MEC Program ``EUROQUAM" (``Fermix").


\begin{thebibliography}{cc}
\bibitem{note.density} A closer inspection into density correlations shows that,
due to the attractive interactions,
each bound pair is actually surrounded by a weak cloud
of excess $\uparrow$ atoms, and viceversa the density-density
correlations of $\downarrow$ atoms shows a weak modulation
at the Fermi wavevector of the $\uparrow$ ones.
Yet these aspects turn out to be irrelevant for the
dominant behavior of spin-spin correlations.
\end{thebibliography}
\end{document}